\documentclass[reqno]{amsart}

\usepackage[utf8]{inputenc}

\usepackage{mathtools}
\mathtoolsset{showonlyrefs}

\usepackage{microtype}
\usepackage[T1]{fontenc}
\usepackage{mathrsfs}
\usepackage{mathptmx}

\usepackage{multirow}
\usepackage[table]{xcolor}
\newcommand{\hl}{\cellcolor{cyan!20}}

\usepackage{algorithm}
\usepackage[noend]{algpseudocode}

\usepackage{amsthm}

\newtheorem{proposition}{Proposition}

\newtheorem{lemma}{Lemma}

\newcommand{\QMC}{\operatorname{QMC}}
\newcommand{\MC}{\operatorname{MC}}
\newcommand{\Var}{\operatorname{Var}}
\newcommand{\R}{\mathbb{R}}
\newcommand{\N}{\mathbb{N}}
\newcommand{\E}{\mathbf{E}}
\newcommand{\1}{\mathbf{1}}
\renewcommand{\P}{\mathbf{P}}
\newcommand{\sF}{\mathscr{F}}
\newcommand{\QV}[1]{\langle#1\rangle}
\newcommand{\STATE}{\State}

\newcommand{\REQUIRE}{\Require}
\newcommand{\FOR}{\For}
\newcommand{\ENDFOR}{\EndFor}

\renewcommand{\epsilon}{\varepsilon}
\newcommand{\bu}{\mathbf{u}}
\newcommand{\bx}{\mathbf{x}}
\newcommand{\by}{\mathbf{y}}
\newcommand{\vxi}{\mathbf{\xi}}
\renewcommand{\Re}{\mathfrak{R}}
\newcommand{\Poisson}{\operatorname{Poisson}}
\newcommand{\Bessel}{\operatorname{Bessel}}
\newcommand{\Gam}{\operatorname{Gamma}}

\begin{document}
\title{Quasi-Monte Carlo methods for the Heston model}
\author{Jan Baldeaux}
\address{Finance Discipline Group\\ University of Technology, Sydney\\ PO Box 123, Broadway, NSW 2007\\ Australia}
\author{Dale Roberts}
\address{Mathematical Sciences Institute and Research School of Finance, Actuarial Studies, and Applied Statistics\\
Australian National University\\
ACT 0200\\
Australia}

\begin{abstract}
In this paper, we discuss the application of quasi-Monte Carlo methods to the Heston model. We base our algorithms on the Broadie-Kaya algorithm, an exact simulation scheme for the Heston model. As the joint transition densities are not available in closed-form, the Linear Transformation method due to Imai and Tan, a popular and widely applicable method to improve the effectiveness of quasi-Monte Carlo methods, cannot be employed in the context of path-dependent options when the underlying price process follows the Heston model. Consequently, we tailor quasi-Monte Carlo methods directly to the Heston model. The contributions of the paper are threefold: We firstly show how to apply quasi-Monte Carlo methods in the context of the Heston model and the SVJ model, secondly that quasi-Monte Carlo methods improve on Monte Carlo methods, and thirdly how to improve the effectiveness of quasi-Monte Carlo methods by using bridge constructions tailored to the Heston and SVJ models. Finally, we provide some extensions for computing greeks, barrier options, multidimensional and multi-asset pricing, and the $3/2$ model.
\end{abstract}

\maketitle

\allowdisplaybreaks

\section{Introduction} \label{secintro}

In this paper, we show how to apply quasi-Monte Carlo (QMC) methods to price path-dependent contingent claims where the underlying asset price is given by the Heston model \cite{Heston93} and the SVJ model \cite{Bates96}. We recall that, on a filtered probability space $(\Omega, \sF, (\sF_t)_{t \ge 0} , \P)$ under the assumption that $\P$ is (already) the risk-neutral pricing measure, the Heston model is given by the system of stochastic differential equations
\begin{equation}
\label{eq:Heston}
\begin{aligned}
dS_t &= r S_t dt + \sqrt{V_t} S_t dW_t^1,\\
dV_t &= \kappa (\theta - V_t) dt + \sigma \sqrt{V_t} dW^2_t,
\end{aligned}
\end{equation}
where $(W^1_t)_{t \ge 0}$ and $(W^2_t)_{t \ge 0}$ are Brownian motions under $\P$ with $d\QV{W^1,W^2} = \rho dt$. The process $S:=(S_t)_{t \ge 0}$ models the asset price dynamics and $V:=(V_t)_{t \ge 0}$ the (stochastic) variance of $S$. Here, $r$ is the risk-free rate of interest, $\theta$ is the long-term average variance, $\kappa$ is the mean-reversion speed of $V$, and $\sigma$ is the volatility of $V$. The SVJ model adds jumps in the dynamics of the underlying $S$ and provides a model that calibrates better to the observed market prices of short-dated European options exhibiting steep skew, see Section~\ref{secExtensionSVJmodel}. An important feature of our approach is that we can work in regimes where the Feller condition $2 \kappa \theta \ge \sigma^2$ is violated, which is useful for market practitioners, see e.g. \cite{DaFGras10}.

The Heston model assumes diffusion dynamics for both the spot price and the volatility, but even though the spot price and volatility are jointly Markovian, we do not deal with a stationary, independent increment process. Therefore our situation is different to the majority of papers that apply QMC methods to finance problems \cite{AlbrecherPr04,AvramidisEfficient,Bal08,B08a,BenthGrKe06,CaflishMoOw97,HartingerSimulation,ImaiTa07,ImaiTa09,Leobacher06,RibeiroValuing,WangSl11} and, in particular, the Linear Transform (LT) method of Imai and Tan \cite{ImaiTa07, ImaiTa09} (see also Leobacher \cite{Leobacher06}) does not seem applicable in our case of path-dependent options when the underlying follows either the Heston model or the SVJ model.

We recall that the LT method relies on the observation that multiplying a vector of standard normal random variates by an orthogonal matrix produces another vector of standard normal random variates. Consequently, evaluating the normal cumulative distribution function elementwise at the vector resulting from the multiplication of an orthogonal matrix by a vector of standard normal random variates, produces a vector of random variates uniformly distributed on $[0,1]$. One consequently uses these uniform random variates in the QMC procedure relevant to the problem. The key step of the LT method is a judicious choice of the orthogonal matrix. In fact, an optimization procedure is used to obtain this matrix, see \cite{ImaiTa07}, \cite{ImaiTa09}. Unfortunately, in \cite{ImaiTa09} the method is only presented for independent increment processes (see Proposition 4.1 in \cite{ImaiTa09}) and requires knowledge of the transition density of the underlying stochastic processes. Though the density of the spot price, conditional on the initial values of spot and volatility, in the Heston model has been studied \cite{BanoRollinFeUt10}, the joint transition density of spot and volatility, which is needed for the valuation of path-dependent options, does not seem to be available. Consequently, the LT method is not applicable in our situation, instead we directly construct bridges for the stock price process under consideration.

We recall that QMC is a class of numerical methods for high-dimensional integrals that can broadly be divided into two categories: nets \cite{DP09,N92} and lattice rules \cite{N92,SJ94}. However, in practice, once the problem is properly formulated both approaches may be applied. This is the first of three contributions of the paper: we first show how to formulate the finance problem as an integration problem based on the exact simulation scheme of Broadie and Kaya \cite{BroadieKa06} then, similar to the results from the references cited above, we demonstrate that QMC methods outperform Monte Carlo (MC) methods when using Sobol point sets with Owen's scrambling method. We also improve our QMC results slightly by conditioning in the case of European call options.

Secondly, we extend the results for the Heston and SVJ model to the case of path-dependent options and demonstrate that by allocating more of the variance to the early dimensions of the QMC point set, the performance of QMC methods can be significantly improved. This is also in line with the results presented in the cited references. We achieve this by employing a bridge construction for the variance process based on \cite{MakarovGl10,YuanKa00} and the well-known bridge construction for a Brownian motion with drift and time-dependent volatility \cite{Glasserman04} and the bridge construction for jump processes \cite{Bal08}. These constructions are in the spirit of \cite{AvramidisEfficient,Bal08,RibeiroValuing}, where bridge constructions were derived for the stochastic processes under consideration. We demonstrate our bridge algorithm by considering Asian call options in the case of the Heston and SVJ model.

Thirdly, we provide a number of additional results of interest in finance. We show how to compute greeks in our framework, we consider Barrier options using the ideas of Glasserman and Staum \cite{GlassermanSt01}, multi-asset stochastic volatility models, and the extension to the $3/2$ model.

The remainder of the paper is structured as follows. In Section~\ref{secBKAlgorithm}, we discuss the Broadie-Kaya algorithm, which forms the basis of the algorithms introduced in this paper, and show how to combine it with the QMC methods. In Section \ref{secQMC}, we recall QMC methods and Section \ref{secEurcalloptions} shows how to effectively combine the Broadie-Kaya algorithm with QMC methods in the context of path-independent European options. Section \ref{secBridgesamplingBK} introduces bridge sampling for the square-root process and the spot price process. The extension to the SVJ model is discussed in Section \ref{secExtensionSVJmodel}. Further extensions of the algorithms presented in this paper are discussed in Section \ref{secFurExt}.

\section{Quantile function for the Heston model} \label{secBKAlgorithm}

To price any contingent claim using a MC or QMC method, one must be able to sample from the law of the asset price process in a fast and accurate way. Our approach is to apply the inverse transform method which requires us to obtain the quantile function of the law of the Heston process $(S_t)_{t \ge 0}$. We recall that given a random variable $X$ with distribution function $F(x) := \P(X \le x)$, the \emph{quantile function} (or \emph{inverse cumulative distribution function}) returns the value below which random samples from the given distribution would occur $p$ amount of the time: the quantile function $Q:[0,1]\to\R$ for the distribution of $X$ is defined as $Q_X(p) := \inf\{x \in \R: p \le F(x)\}$.

For $t > u$, the distribution of $S_t$ and $V_t$ solving \eqref{eq:Heston} with initial conditions $S_u \in \R_+$ and $V_u \in \R_+$ can easily be shown to be given by
\begin{align}
 S_t &= S_u \exp \left( r (t - u) - \frac{1}{2} \int^t_u V_s ds + \rho \int^t_u \sqrt{V_s} dW^2_s + \sqrt{1-\rho^2}\int_u^t \sqrt{V_s}\,dW_s^1 \right),\nonumber\\
V_t &= V_u + \kappa \theta (t-s) - \kappa \int^t_s V_u du + \sigma \int^t_s \sqrt{V_u} dW^2_u. \label{eqvarHeston}
\end{align}
Our approach to obtain the quantile function for the distribution of the random variable $S_t$ is based on the exact simulation method obtained by Broadie and Kaya \cite{BroadieKa06}.

\subsection{The Broadie-Kaya approach}

We recall that the exact simulation approach for \eqref{eq:Heston} given by Broadie and Kaya \cite{BroadieKa06} is as follows:
\begin{enumerate}
\item Simulate $V_t$ given $V_u = x$,
\item Generate a sample from the distribution of $\int^t_u V_s ds$ given $V_u = x$ and $V_t = y$,
\item Recover $\int^t_u \sqrt{V_s} dW^2_s$ from \eqref{eqvarHeston} given $V_t = y$, $V_u = x$, and $\int^t_u V_s ds = z$ as
\[ \int^t_s \sqrt{V_u} dW^2_u = \frac{1}{\sigma} \left(y - x + \kappa \theta (t-u) - z\right), \]
\item Generate a sample from the distribution of $S_t$ given $\int_u^t \sqrt{V_s}dW_s^2$ and $\int_u^t V_s\,ds$.
\end{enumerate}

When simulating a random variable $X$ using QMC methods it is sometimes more convenient to rephrase such algorithm in terms of quantiles, allowing one to substitute either random uniforms (for a MC method) or QMC point sets in the quantile function $Q_X$ to draw a sample of $X$.

\subsection{Obtaining quantile functions}

The Broadie-Kaya approach can be reformulated in terms of quantiles.  To simplify notation, we shall henceforth use the following convention: we write $X \sim L$ to denote that $X$ has a law $L$ and $X|Y=y \sim L$ to denote that the random variable $X$ conditioned on the event $Y=y$ has law $L$. We sometimes write $X|y$ instead of $X|Y=y$ when the context is clear.

The quantile function for $V_t|V_u = x$, which is needed for step 1 in the Broadie Kaya approach, is easily found. It is well known that for $t > u$ the distribution of $V_t | V_u$ follows a noncentral chi-squared distribution:
\begin{displaymath}
V_t | V_u = x \sim \frac{\sigma^2 ( 1- \exp ( - \kappa ( t - u) ) )}{4 \kappa } \chi^2_d \left( \frac{4 \kappa \exp (- \kappa ( t - u)) }{\sigma^2 (1 - \exp ( - \kappa (t - u) ) )} x \right), \quad d := \frac{4 \theta \kappa}{\sigma^2},
\end{displaymath}
where $\chi^2_\nu (\lambda)$ denotes the noncentral chi-squared random variable with $\nu$ degrees of freedom and noncentrality parameter $\lambda$. Therefore, the quantile function for $V_t | V_u = x$ can be obtained from the quantile function for a noncentral chi-squared distribution with the appropriate choice of parameters $\nu$ and $\lambda$.

The quantile function of $\int_u^t V_s\,ds|Vu=x, V_t = y$ is the most expensive step. Broadie and Kaya obtained its distribution function by computing the conditional Laplace transform and performing a numerical inversion of the characteristic function. An alternative approach to determining the distribution function was given by Glasserman and Kim using a series expansion of the random variable \cite{GlassermanKi11}. We follow the approach given by Broadie and Kaya whereby the conditional distribution $F(x) :=\P\left(\int_u^t V_s\,ds \le x |V_u,V_t\right)$ is obtained by numerically inverting the (conditional) characteristic function $\Phi(a) := \E\left[\exp\left(i a \int_u^t V_s\,ds \bigl| V_u,V_t\bigr.\right)\right]$. We recall that $\Phi$ was given explicitly as
\begin{displaymath}
\Phi(a) = \frac{\gamma(a) e^{-\frac12(\gamma(a)-\kappa)(t-u)}(1-e^{-\kappa(t-u))})}{\kappa(1-e^{-\gamma(a)(t-u)})} \frac{A(a)}{B}\exp(C(a))\\
\end{displaymath}
where the terms $A$, $B$, $C$, and $\gamma(a)$ are
\begin{align*}
A (a)&:= I_{d/2-1} \left(\sqrt{V_u V_t} \frac{4\gamma(a) e^{\frac12 \gamma(a)(t-u)}}{\sigma^2(1-e^{-\gamma(a)(t-u)})}\right), \\
B &:= I_{d/2-1} \left(\sqrt{V_u V_t} \frac{4 \kappa e^{\frac12 \kappa (t-u)}}{\sigma^2(1-e^{-\kappa(t-u)})}\right),\\
C (a)&:= \frac{V_u + V_t}{\sigma^2} \left(\frac{\kappa(1+e^{\kappa(t-u)})}{1-e^{-\kappa(t-u)}} - \frac{\gamma(a)(1+e^{-\gamma(a)(t-u)})}{1-e^{\gamma(a)(t-u)}}\right),\\
\gamma(a) &:= \sqrt{\kappa^2 - 2 \sigma^2 i a},
\end{align*}
and $I_\nu(x)$ denotes the modified Bessel function of the first kind.
The inversion of $\Phi$ (which gives $F(x)$), can be calculated by applying a trapezoidal rule to approximate the integral
\[
	F(x) = \frac{2}{\pi} \int_0^\infty \frac{\sin(zx)}{z} \Re \Phi(z)\,dz,
\]
where $\Re(x+iy) := x$ for $x,y \in \R$. We note that the term $B$ does not depend on $a$ so only needs to be computed once per inversion of $\Phi$. Given a mesh size of $h$, the integral is approximated as
\begin{equation}
	F(x) = \frac{hx}{\pi} + \frac{2}{\pi} \sum_{j=1}^N \frac{\sin(hjx)}{j} \Re \Phi(hj) - \epsilon_h - \epsilon_N
\end{equation}
where $\epsilon_h$ is the discretization error associated with the choice of mesh size $h$ and $\epsilon_N$ is the truncation error caused by taking $N < \infty$. Following Broadie and Kaya, to approximate $F(x)$ with good accuracy we set $h = 2 \pi / (x+|m_1| + q |\sqrt{m_2 - m_1^2}|)$ where $m_1$ and $m_2$ are the first and second moments of $\int_u^t V_s\,ds|V_u,V_t$ and $q \in \N$ is chosen sufficiently large (e.g., $q=5$). Explicit (but long) expressions for $m_1$ and $m_2$ can be found using a computer algebra system (CAS) using the well-known technique of differentiating $\Phi$ and setting $a=0$. The upper bound of the summation is chosen to satisfy $|\Phi(hN)|/N < \pi \epsilon / 2$ where $\epsilon$ is the desired truncation error. Thus, given our numerical approximation $F^{h,N}$ of the distribution function $F$ we can apply a root finding procedure to identify the quantile function of $F^{h,N}$.

Finally, we consider the quantile function of $S_t$ conditioned on the event $\int_u^t V_s\,ds = y$, the event $\int_u^t \sqrt{V_s}dW_s^2 = z$, and the initial condition $S_u$. First, observe that $\sqrt{1-\rho^2} \int_u^t \sqrt{V_s}\,dW_s^1$ is normally distributed with mean zero and variance $\int_u^t V_s\,ds$. Therefore, $\log(S_t)|y,z,S_u$ is normally distributed with mean $r(t-u)-\frac{1}{2} y + \rho z$ and variance $(1-\rho^2) \int_u^tV_s\,ds = (1-\rho^2) y$. Further, if we have the quantile function $Q_Z$ for $Z \sim N(0,1)$ then if $X \sim N(\mu,\sigma^2)$ then $Q_X(p) = \mu + \sigma Q_Z(p)$ and $Q_{\exp(X)}(p) = \exp(\mu + \sigma Q_Z(p))$ as $e^x$ is an increasing function of $x$. Hence, we have the quantile function for $S_t|y,z,S_u$.


\section{QMC methods} \label{secQMC}

Regarding QMC point sets, we will make use of digital nets \cite{DP09,N92}. Given a (base $b$) digital net $(\bu_i)^n_{i=1}$, where $\bu_i \in [0,1]^d$ for some dimension $d \in \mathbb{Z}^+$, we will always randomize the point set using Owen's scrambling algorithm to compute standard errors \cite{O95}. Given a generic point $\bx \in [0,1)^d$, where $\bx=(x_1, \dots, x_s)$, we recall that Owen's algorithm expands each $x_j$ as
\begin{displaymath}
x_j = \frac{\xi_{j,1}}{b} + \frac{\xi_{j,2}}{b^2} + \dots \, ,
\end{displaymath}
and generates a scrambled point $\by \in [0,1)^d$, where $\by=(y_1, \dots, y_d)$, as
\begin{displaymath}
y_j = \frac{\eta_{j,1}}{b} + \frac{\eta_{j,2}}{b^2} + \dots \, .
\end{displaymath}
The permutation $\pi_j$ applied to $\xi_{j,l}$, $j=1,\dots,d$ depends on $\xi_{j,k}$, for $1 \leq l$. In particular, $\eta_{j,1}=\pi_j( \xi_{j,1})$, $\eta_{j,2} = \pi_{j, \xi_{j,1}} (\xi_{j,2})$, $\eta_{j,3} = \pi_{j, \xi_{j,1}, \xi_{j,2}} (\xi_{j,3})$ and in general
\begin{displaymath}
\eta_{j,k} = \pi_{j, \xi_{j,1}, \dots, \xi_{j,k-1}} (\xi_{j,k}) \, , \, k \geq 2 \, ,
\end{displaymath}
where $\pi_j$ and $\pi_{j, \xi_{j,1}, \dots, \xi_{j,k-1}}$, $k \geq 2$, are random permutations of $\left\{ 0,1 , \dots, b-1 \right\}$. We assume that permutations with different indices are mutually independent. We recall that if the scrambling algorithm is applied to $\bx$ to obtain $\by$, then $\by$ is uniformly distributed in $[0,1)^d$ by Proposition 2 in \cite{O95}. We find it convenient to introduce the following notation
\begin{displaymath}
\by = \pi(\bx) \, ,
\end{displaymath}
so $\pi( \cdot)$ represents the scrambling algorithm applied to $\bx$ to obtain $\by$.

We recall that by our reformulation of the Broadie Kaya approach for the Heston model in Section~\ref{secBKAlgorithm} in terms of quantiles, the discounted payoff $e^{-rT} g(S_T)$ for some $T > 0$ can be rewritten as a function $f:[0,1]^3 \to \R$ and, in the case of a path-dependent payoff, $e^{-rT} g(S_{t_1}, S_{t_2}, \ldots, S_{t_n})$ for times $0 < t_1, t_2, \ldots, t_n = T$ can be transformed to the function $f:[0,1]^{3 \times n} \to \R$. Henceforth, we simply assume that all discounted payoffs are mapped to the $d$-dimensional (for some appropriate $d$) unit cube in this manner.

Given $q$ independent permutations $\pi^r$, $r=1,\dots,q$, and the discounted payoff of the financial derivative $f:[0,1]^d \rightarrow \R$, we estimate the price of the derivative using
\begin{displaymath}
I_{\QMC} = \frac{1}{q} \sum^q_{r=1} I_r = \frac{1}{q} \sum^q_{r=1} \frac{1}{n} \sum^n_{i=1} f( \pi^r ( \bu_i)),
\end{displaymath}
and compute standard errors via
\begin{displaymath}
\sigma_{\QMC} = \sqrt{ \frac{ \sum^q_{r=1} ( I_r - I_{\QMC})^2 }{q(q-1)} } \, .
\end{displaymath}
We point out that for digital nets, scrambling is the preferred randomization method, as it achieves the optimal convergence rate \cite{BD10,DP09,O97}. For an implementation of the scrambling algorithm, we refer the reader to \cite{HongHi03}.

For purposes of comparison, we will also look at MC estimators. In this case, we will choose $q \times n$ points, $(\vxi_i)^{q \times n}_{i=1}$, independent and identically distributed in $[0,1]^d$, and estimate derivative prices using
\begin{displaymath}
I_{\MC} = \frac{1}{q \times n} \sum^{q \times n}_{i=1} f ( \vxi_i)
\end{displaymath}
and compute standard errors using
\begin{displaymath}
\sigma_{\MC} = \sqrt{ \frac{\sum^{q \times n}_{i=1} ( f ( \vxi_i) - I_{\MC})^2}{q \times n ( q \times n -1)} } \, .
\end{displaymath}
We conclude the section by commenting on the variances of $I_{\MC}$ and $I_{\QMC}$. For square-integrable functions $f$, it is well-known that
\begin{displaymath}
\Var(I_{\MC}) = \frac{\Var(f(\vxi_i))}{q \times n} \, .
\end{displaymath}
Regarding $I_{\QMC}$, it is known that for a square-integrable function $f$ the Monte Carlo convergence rate is matched \cite{O97a,DP09}. However, scrambled digital nets can exploit the smoothness of the integrand: if the integrand satisfies a H\"older condition of order $\alpha$ with $0 < \alpha \leq 1$, then the variance decays at a rate of order $\mathcal{O}( n^{-(1+ 2 \alpha) + \varepsilon})$ for some $\varepsilon >0$, and the leading constant is allowed to depend on the dimension of the point set \cite{BD10,DP09}. In practise, it is difficult to confirm the smoothness of the integrand under consideration, hence it is important to investigate the standard errors of QMC methods numerically.

\section{European Call options} \label{secEurcalloptions}

In this section, we follow \cite{Willard97} and apply conditional Monte Carlo method to improve the efficiency of the Broadie-Kaya algorithm. We use European call options to demonstrate the method. The approach relies on the observation that given 
\begin{displaymath}
\int^T_0 V_s ds \, , \, \int^T_0 \sqrt{V_s} d W^2_s
\end{displaymath}
the price of a European call option is given by the Black-Scholes price, with modified initial share price
\begin{displaymath}
\tilde{S}_0 = S_0 \exp \left( - \frac{\rho^2}{2} \int^T_0 V_s ds +\rho \int^T_0 \sqrt{V_s} dW^2_s \right)
\end{displaymath}
and adjusted volatility $\tilde{\sigma} \sqrt{1 - \rho^2}$, where $\tilde{\sigma} := \sqrt{\frac{1}{T} \int^T_0 V_s ds}$. Using $BS(S_0, K, r, \tau, \sigma)$ to denote the Black-Scholes price of a European call with initial stock price $S_0$, strike $K$, interest rate $r$, time to maturity $\tau$ and volatility $\sigma$, we have
\begin{displaymath}
E \left( e^{ - r T} \left( S_T - K \right)^+ \right) = E \left(  BS \left( \tilde{S}_0, K , r , \tau , \tilde{\sigma} \right) \right)  \, ,
\end{displaymath}
that is, we firstly simulate $\int^T_0 V_s ds$ and $\int^T_0 \sqrt{V_s} d W^2_s$ using the Broadie-Kaya algorithm, and then compute the Black-Scholes price, for the particular values of $\tilde{S}_0$ and $\tilde{\sigma}$ corresponding to $\int^T_0 V_s ds$ and $\int^T_0 \sqrt{V_s} d W^2_s$. In Table \ref{tabEuropeans}, we show price and standard error estimates for a European call option. Columns 1 and 2 show that combining QMC methods with the Broadie-Kaya algorithm already improves on the Monte Carlo method. However, using the conditioning argument, the QMC method significantly improves on a naive application of QMC methods to the Broadie-Kaya algorithm. There are two reasons for the variance reduction: we do not estimate the Black-Scholes price using Monte Carlo simulation, which is done in Step 4 of the Broadie-Kaya algorithm, but compute the value exactly. Furthermore, it can be expected that when combining the conditional expectations with QMC methods, the approach is even more efficient \cite{Willard97}. This is due to the fact that taking the conditional expectation has a smoothing effect, which can be expected to improve the performance of quasi-Monte Carlo methods, see \cite[Subsection 10.1]{LEcuyerLe00}.

\begin{table}
\caption{Estimated prices (with standard errors given in parentheses) of European options where the asset price process is given by a Heston model. These values are based on 30 independent batches.} \label{tabEuropeans}
\begin{center}
\begin{tabular}{ccccc} 
	\hline
	Trials & MC & QMC & Cond QMC \\
	\hline
         128 &           6.847642 (0.121792) &          6.792353 (0.074884) &          6.807731 (0.011452) \\
         256 &           6.785936 (0.085627) &          6.815575 (0.037468) &          6.807578 (0.007037) \\
         512 &           6.794658 (0.059212) &          6.807123 (0.023326) &          6.806918 (0.001924) \\
        1024 &           6.818196 (0.042269) &          6.805009 (0.010747) &          6.807080 (0.001928) \\
        2048 &           6.820823 (0.030005) &          6.805464 (0.004077) &          6.806219 (0.001182) \\
        4096 &           6.815857 (0.021199) &          6.806346 (0.002346) &          6.806326 (0.000480) \\
        8192 &           6.789039 (0.014945) &          6.806315 (0.001721) &          6.806558 (0.000520) \\
       16384 &           6.800179 (0.010576) &          6.806484 (0.000730) &          6.806438 (0.000249) \\
   \hline  
\end{tabular}
\\[0.3em]
{\small Option parameters: $S=100$, $K=100$, $V_0=0.010201$, $\kappa=6.21$, $\theta=0.019$, $\sigma=0.61$, $\rho=-0.70$, $r=3.19\%$, $T=1.0$ year, true option price is $6.80611$.}
\end{center}
\end{table}

\section{Bridge sampling for path-dependent options} \label{secBridgesamplingBK}

In this section, we study the pricing of options whose payoff is a function of
\begin{displaymath}
S_{t_1} , \ldots , S_{t_h} ,
\end{displaymath}
where we choose $h=2^m$ for simplicity. A naive approach to simulating this path is given in Algorithm~\ref{Naive}. The algorithms for the simulation of the random variables required in steps 1, 2, and 3 of Algorithm~\ref{Naive} (using either MC or QMC) follows from the algorithm given in Section~\ref{secBKAlgorithm}.

\begin{algorithm}
\caption{Naive version of the Broadie-Kaya exact simulation algorithm} \label{Naive}
\begin{algorithmic}[1]
\STATE \label{Naive1} Simulate $( V_{t_i} )^{N}_{i=1}$ in the following order
\begin{displaymath}
V_{t_1} \, , \, V_{t_2} \, , \, \cdots \, , \,  V_{t_N} \, .
\end{displaymath}
\STATE Simulate $ \left( \int^{t_i}_{t_{i-1}} V_s ds \vert V_{t_{i-1}} , V_{t_i} \right)$ for $i=1, \dots , N$.
\STATE Simulate $(S_{t_i})^{N}_{i=1}$ in the following order
\begin{displaymath}
S_{t_1} \, , \, S_{t_2} \, , \, S_{t_3} \, , \, \cdots \, , S_{t_N} \, .
\end{displaymath}
\end{algorithmic}
\end{algorithm}

It is well-known that QMC methods are particularly efficient if the effective dimension of the problem under consideration is low \cite{CaflishMoOw97}. In our case, the effective dimension depends on the variance of the discounted pay-off of the financial derivative under consideration. To reduce the effective dimension, following \cite{AvramidisEfficient,Bal08,CaflishMoOw97}, we now allocate the early dimensions to variates with high variances. We achieve this by proposing a bridge algorithm, given by Algorithm \ref{Bridge}, to generate the paths of $(S_t)_{t \ge 0}$  at the time points $t_1, \ldots, t_N$. The simulation of the random variables required in our bridge algorithm requires the construction of a bridge sampling algorithm for the square-root process $(V_t)_{t \ge 0}$ and a bridge sampling algorithm for the stock price process $(S_t)_{t \ge 0}$. In the coming sections, we propose algorithms to construct these bridges.

\begin{algorithm}
\caption{Bridge version of the exact simulation algorithm}\label{Bridge}
\begin{algorithmic}[1]\label{algbridgeBKalg1}
\STATE Simulate $( V_{t_i} )^{N}_{i=1}$ in the following order
\begin{displaymath}
V_{t_N} \, , \, V_{t_{\frac{N}{2}}} \, , \,  V_{t_{\frac{N}{4}}} \, , \,  V_{t_{\frac{3 N}{4}}} \, , \cdots
\end{displaymath}
\STATE Simulate $ \left( \int^{t_i}_{t_{i-1}} V_s ds \vert V_{t_{i-1}} , V_{t_i} \right)$ for $i=1, \dots , N$.
\STATE Simulate $(S_{t_i})^{N}_{i=1}$ in the following order
\begin{displaymath}
S_{t_N} \, , \, S_{t_{\frac{N}{2}}} \, , \, S_{t_{\frac{N}{4}}} \, , \, S_{t_{\frac{3N}{4}}} \, , \, \dots
\end{displaymath}
\end{algorithmic}
\end{algorithm}

\subsection{Bridge sampling for square-root processes} \label{subsecbridgesamplingSRprocess}

We now recall how to perform bridge sampling for square-root processes by relying on the bridge sampling algorithm for squared Bessel processes studied by Yuan and Kalbfleisch \cite{YuanKa00} and Makarov and Glew \cite{MakarovGl10}. The following well-known result linking square-root processes and squared Bessel processes is employed, see Proposition 6.3.1.1 in \cite{JeanblancYoCh09}.

\begin{proposition} The square-root process $V= \left\{ V_t \, , \, t \geq 0 \right\}$ is a squared Bessel process transformed by the following space-time change:
\begin{displaymath}
V_t = e^{- \kappa t } X_{c(t)} \, ,
\end{displaymath}
where $c(t) = \frac{\sigma^2}{4 \kappa} ( e^{\kappa t } - 1)$, $X=\left\{ X_t \, , \, t \geq 0  \right\}$ is a squared Bessel process of dimension $\delta = \frac{ 4 \kappa \theta }{\sigma^2}$.
\end{proposition}
As we assume that $\kappa, \theta, \sigma >0$, we have that for $\delta \geq 2$, the boundary $0$ is polar and for $0< \delta < 2$ it is reflecting, see Figure 6.1 in \cite{JeanblancYoCh09}. This allows us to propose Algorithm~\ref{algbridgeBK}, adapted from the algorithm in \cite{MakarovGl10}, to sample a square-root process at times $t_1, \ldots,  t_h$ where $h=2^m$ with $m \geq 1$.

\begin{algorithm}
\caption{Bridge sampling for the square-root process} \label{algbridgeBK}
\begin{algorithmic}[1]\label{algbridgesr}
\REQUIRE Time indices $(t_1 , \dots, t_{2^m})$
\STATE $h \leftarrow 2^m$, $j_{\max} \leftarrow 1$, $x_0 \leftarrow V_0$, $\delta \leftarrow \frac{4 \kappa \theta}{\sigma^2}$
\FOR{$i=1, \ldots, h$}
$s_i \leftarrow \frac{\sigma^2}{4 \kappa} ( \exp ( \kappa t_i )  - 1)$
\ENDFOR
\STATE Generate $Z \sim \chi^2(\delta, x_0/s_N)$
\STATE $x_h \leftarrow s_h Z$
\STATE $t_0 \leftarrow 0$
\FOR{$k=1, \dots, m$}
\STATE $i_{\min} \leftarrow h/2$, $i \leftarrow i_{\min}$
\STATE $l \leftarrow 0$, $r \leftarrow h$
\FOR{$j=1, \dots, j_{\max}$}
\STATE $\lambda \leftarrow \frac{1}{2 (t_r - t_l)} \left( \frac{t_r - t_i}{t_i - t_l} x_l + \frac{t_i - t_l}{t_r - t_i} x_r \right)$
\STATE Generate $P \sim \Poisson( \lambda)$
\STATE Generate $Z \sim \Bessel(\frac{\delta}{2} -1, \frac{ \sqrt{x_l x_r}}{t_r - t_l} )$
\STATE Generate $G \sim \Gam( P + 2 Z + \frac{\delta}{2} , \frac{t_r - t_l}{2 (t_i - t_l)(t_r - t_i)})$
\STATE $x_i \leftarrow G$
\STATE $i \leftarrow i +h$, $l \leftarrow l +h$, $r \gets r +h$
\ENDFOR
\STATE $j_{\max} \leftarrow 2 j_{\max}$, $h \leftarrow i_{\min}$
\ENDFOR
\FOR{$i=1, \ldots, h$}
$v_i \leftarrow \exp ( - \kappa t_i) x_i$
\ENDFOR
\Return sampled path $(v_1 , \dots, v_{h})$ from distribution of $(V_{t_1} , \dots, V_{t_{2^m}} )$
\end{algorithmic}
\end{algorithm}

\subsection{Bridge sampling for the stock price} \label{subsecbridgesamplingstockprice}

We now propose a bridge sampling algorithm for step 3 of Algorithm~\ref{Bridge}, that is, an algorithm to simulate $(S_{t_i})_{i=1}^h$ where $h=2^m$ in the order $S_{t_h}, S_{t_{h/2}}, S_{t_{h/4}}, S_{t_{3h/4}}, \ldots$ with effect of allocating the early dimensions to variates with high variances. The following lemma allows us to set up a sampling scheme for the stock price process.

\begin{lemma} \label{lemjointdistrstock} Given times $u < t_1 < t_2 < t_3$, the joint distribution of $\log(S_{t_1}), \log(S_{t_2}), \log(S_{t_3})$ conditional on the initial condition $S_u$, the variances $V:=(V_{t_1}, V_{t_2}, V_{t_3})$, and the integrated variances $IV := ( \int^{t_1}_u V_s\,ds , \int^{t_2}_u V_s\,ds , \int^{t_3}_u V_s\,ds )$ is given by
\begin{displaymath}
\left. \left( \begin{array}{c} \log (S_{t_1}) \\ \log(S_{t_2}) \\ \log(S_{t_3}) \end{array} \right) \right| V, IV, S_u \sim N \left(  \left[ \begin{array}{c} \mu_1 \\ \mu_2 \\ \mu_3 \end{array} \right] , \left[  \begin{array}{ccc} \sigma^2_1 & \sigma^2_1 & \sigma^2_1 \\ \sigma^2_1 & \sigma^2_2 & \sigma^2_2 \\ \sigma^2_1 & \sigma^2_2 & \sigma^2_3 \end{array} \right]     \right),
\end{displaymath}
where for $i=1,2,3$ we have defined $\sigma^2_i := (1 - \rho^2) \int^{t_i}_u V_s\,ds$ and
\begin{displaymath}
\mu_i := \log(S_{t_0} ) + r t_i - \frac{1}{2} \int^{t_i}_u V_s\, ds + \rho \int^{t_i}_u \sqrt{V_s}\,dW^1_s.
\end{displaymath}
Further, we have $\log(S_{t_2}) | S_{t_1}, S_{t_3}, V, IV \sim N(M, \Sigma)$
where
\begin{equation*}
	M:=  \log(S_{t_1}) + \mu_2 - \mu_1 + \frac{\sigma^2(t_2) - \sigma^2(t_1)}{\sigma^2(t_3) - \sigma^2(t_1)} \left( \log(S_{t_3}) - \log(S_{t_1}) + \mu_1 - \mu_3\right),
\end{equation*}
and
\begin{equation*}
	\Sigma :=\sigma^2(t_2) - \sigma^2(t_1) - \frac{(\sigma^2(t_2) - \sigma^2(t_1) )^2 }{\sigma^2(t_3) - \sigma^2(t_1)}.
\end{equation*}
\end{lemma}

\begin{proof} The first part of the lemma is straightforward, the second follows immediately from the conditioning formula (2.25) in \cite{Glasserman04}.
\end{proof}

By applying Lemma~\ref{lemjointdistrstock}, we propose Algorithm~\ref{algbridgestock} to sample the path of the share price process $(S_t)_{t \ge 0}$ at the time points $(S_{t_1}, \dots, S_{t_{h}})$. This algorithm requires the \emph{drifts} $(\mu_1, \ldots, \mu_{2^m})$ and \emph{volatilities} $(\sigma_1^2, \ldots, \sigma_{2^m}^2)$ determined in Lemma~\ref{lemjointdistrstock}.

\begin{algorithm}
\caption{Bridge sampling for the share price process}\label{algbridgestock}
\begin{algorithmic}[1]
\REQUIRE Time indices $(t_1 , \dots, t_{2^m})$, drifts $(\mu_1, \dots, \mu_{2^m})$, volatilities $(\sigma^2_1 , \dots, \sigma^2_{2^m})$, and $S_0$
\STATE $h \leftarrow 2^m$, $j_{\max} \leftarrow 1$
\STATE Generate $(Z_1, \dots, Z_{h}) \sim N ( 0 , I)$
\STATE $s_h \leftarrow \exp ( \mu_h + \sigma_h Z_h)$
\STATE $s_0 \leftarrow S_0$
\FOR{$k=1, \dots, m$}
\STATE $i_{\min} \leftarrow h/2$, $i \leftarrow i_{\min}$
\STATE $l \leftarrow 0$, $r \leftarrow h$
\FOR{$j=1, \dots, j_{\max}$}
\STATE $a \leftarrow \mu_i - \mu_l + \log(s_l) + \frac{\sigma^2_i - \sigma^2_l}{\sigma^2_r - \sigma^2_l} \left( \log(s_r) - \log(s_l) + \mu_l - \mu_r \right)$
\STATE $b \leftarrow \sigma^2_i - \sigma^2_l - \frac{( \sigma^2_i - \sigma^2_l )^2}{\sigma^2_r - \sigma^2_l}$
\STATE $s_i \leftarrow \exp( a + b Z_i)$
\STATE $i \leftarrow i +h$, $l \leftarrow l +h$, $r \gets r +h$
\ENDFOR
\STATE $j_{\max} \leftarrow 2 j_{\max}$, $h \leftarrow i_{\min}$
\ENDFOR
\Return sampled path $(s_1 , \dots, s_{2^m})$ from distribution of $(S_{t_1} , \dots, S_{t_{2^m}} )$
\end{algorithmic}
\end{algorithm}

\subsection{Numerical results}

To demonstrate our bridging technique, we estimate the price of an Asian call option where the asset price process is given by a Heston model. We recall that the payoff of an Asian call option, for strike price $K$ and expiry $T$, is given by
\begin{displaymath}
\left( \frac{1}{d} \sum^d_{i=1} S_{t_i} - K \right)^+,
\end{displaymath}
where $t_1, \ldots, t_d$ are $d$ monitoring dates. We compare the prices obtained with standard Monte Carlo, QMC using Sobol points and Owen's scrambling algorithm, and the bridge construction using Sobol points and Owen's scrambling. These results are presented in Table~\ref{TableAsian}. We find that QMC methods already improve on Monte Carlo methods, however, using bridge constructions, the effectiveness of QMC methods can be enhanced.

\begin{table}
{\small
\caption{Standard errors of Asian option price where the asset price process is given by a Heston model using Monte Carlo (M), Quasi-Monte Carlo (Q), and the Bridge Quasi-Monte Carlo (B) algorithms. These values are based on 30 independent batches and $m$ trials.}\label{TableAsian}
\begin{tabular}{c|cccccccccc}
\hline
$n$ & $m$: & $2^6$ & $2^7$ & $2^8$ & $2^9$ & $2^{10}$ & $2^{11}$ & $2^{12}$ & $2^{13}$ & $2^{14}$ \\
\hline
\multirow{3}{*}{$4$} 
& M  & 0.0959 & 0.0740 & 0.0639 & 0.0365 & 0.0302 & 0.0204 & 0.0128 & 0.0086 & 0.0052 \\ 
& Q  & 0.0303 & 0.0231 & 0.0147 & 0.0079 & 0.0039 & 0.0033 & 0.0020 & 0.0009 & 0.0008 \\
& B & \hl 0.0299 & \hl 0.0172 & \hl 0.0133 & \hl 0.0070 & 0.0053 & 0.0036 & 0.0028 & 0.0025 & 0.0010 \\
\hline
\multirow{3}{*}{8}
& M & 0.1054 & 0.0713 & 0.0399 & 0.0338 & 0.0253 & 0.0166 & 0.0126 & 0.0075 & 0.0052 \\
& Q & 0.0413 & 0.0285 & 0.0160 & 0.0080 & 0.0056 & 0.0043 & 0.0023 & 0.0021 & 0.0009 \\
& B & \hl 0.0379 & \hl 0.0280 & \hl 0.0149 & 0.0151 & 0.0063 & \hl 0.0032 & 0.0026 & 0.0021 & 0.0011 \\
\hline
\multirow{3}{*}{16}
& M & 0.0973 & 0.0646 & 0.0435 & 0.0351 & 0.0244 & 0.0227 & 0.0127 & 0.0074 & 0.0069 \\ 
& Q & 0.0462 & 0.0241 & 0.0189 & 0.0102 & 0.0073 & 0.0052 & 0.0027 & 0.0019 & 0.0010 \\
& B & \hl 0.0409 & 0.0298 & \hl 0.0182 & 0.0130 & \hl 0.0061 & \hl 0.0042 & 0.0029 & \hl 0.0018 & 0.0013 \\
\hline
\multirow{3}{*}{32}
& M & 0.0804 & 0.0608 & 0.0432 & 0.0364 & 0.0269 & 0.0163 & 0.0106 & 0.0078 & 0.0055 \\
& Q & 0.0523 & 0.0377 & 0.0181 & 0.0185 & 0.0123 & 0.0095 & 0.0035 & 0.0022 & 0.0018 \\
& B & \hl 0.0339 & \hl 0.0269 & 0.0283 & \hl 0.0114 & \hl 0.0101 & \hl 0.0047 & \hl 0.0030 & 0.0026 & \hl 0.0009 \\
\hline
\multirow{3}{*}{64}
& M & 0.0913 & 0.0656 & 0.0362 & 0.0349 & 0.0246 & 0.0158 & 0.0127 & 0.0080 & 0.0064 \\
& Q & 0.0575 & 0.0400 & 0.0251 & 0.0219 & 0.0182 & 0.0113 & 0.0064 & 0.0034 & 0.0020 \\
& B & \hl 0.0537 & \hl 0.0205 & \hl 0.0118 & \hl 0.0069 & \hl 0.0076 & \hl 0.0061 & \hl 0.0046 & \hl 0.0033 & 0.0039 \\
\hline
\multirow{3}{*}{128}
& M & 0.0920 & 0.0612 & 0.0318 & 0.0314 & 0.0253 & 0.0142 & 0.0113 & 0.0099 & 0.0064 \\ 
& Q & 0.0705 & 0.0331 & 0.0319 & 0.0208 & 0.0201 & 0.0123 & 0.0054 & 0.0038 & 0.0027 \\
& B & \hl 0.0641 & 0.0672 & 0.0587 & \hl 0.0110 & \hl 0.0075 & \hl 0.0066 & \hl 0.0047 & \hl 0.0037 & 0.0034 \\
\hline
\end{tabular}
}
\\[0.3em]
{\small Option parameters: $S=100$, $K=100$, $V_0=0.010201$, $\kappa=6.21$, $\theta=0.019$, $\sigma=0.61$, $\rho=-0.70$, $r=3.19\%$, with $n$ time monitors over the time period $[0,1]$.}
\end{table}

\section{Extension to the SVJ model} \label{secExtensionSVJmodel}

It is well known that Heston stochastic volatility model cannot fit short-term smiles well if they exhibit skew \cite{Gatheral06}. This motivates the introduction of jumps into the dynamics of the underlying share price process. The following model, often called the SVJ model, was first proposed in \cite{Bates96}:
\begin{align}
d S_t  &= S_{t-} \left( (  r - \lambda \bar{\mu} ) dt + V_t \left( \rho dW^1_t + \sqrt{1 - \rho^2} dW^2_t \right) + (Y_t - 1) dN_t \right) \label{eqstockpriceSVJ},
\\ d V_t &= \kappa ( \theta - V_t ) dt + \sigma \sqrt{V_t} dW^1_t , \nonumber
\end{align}
where $N_t$ is a Poisson process with constant intensity $\lambda$, $(W^1_t)_{t \ge 0}$ and $(W^2_t)_{t \ge 0}$ are independent Brownian motions both independent from the Poisson process $(N_t)_{t \ge 0}$, and the jump variables $Y_t$ are a family of independent random variables all having the same lognormal distribution with mean $\mu_S$ and variance $\sigma^2_S$. Furthermore, $\E \left( Y_t -1 \right) = \bar{\mu}$ so it follows that
\begin{displaymath}
\mu_S = \log(1 + \bar{\mu}) - \frac{1}{2} \sigma^2_S.
\end{displaymath}
Solving the SDE for the stock price process \eqref{eqstockpriceSVJ}, we obtain
\begin{equation} \label{eqsplitdiffusionjumpsSVJ}
S_t = \tilde{S}_t \prod^{N_t}_{j=1} \tilde{Y}_j,
\end{equation}
where
\begin{displaymath}
\tilde{S}_t = S_0 \exp \left( ( r - \lambda \bar{\mu} ) t - \frac{1}{2} \int^t_0 V_s\,ds + \rho \int^t_0 \sqrt{V_s}\,dW^1_s + \sqrt{1 - \rho^2} \int^t_0 \sqrt{V_s}\,dW^2_s  \right).
\end{displaymath}

\subsection{Exact simulation for the SVJ model}

As discussed in \cite{BroadieKa06,KornKoKr10}, equation \eqref{eqsplitdiffusionjumpsSVJ} motivates the simulation algorithm for the SVJ model: First simulate the diffusion part as in Section~\ref{secBKAlgorithm} and then take care of the jump part using $\prod^{N_t}_{j=1} \tilde{Y}_j$. This results in Algorithm~\ref{algSVJEurop} which is the analogue of the Broadie Kaya algorithm presented in Section~\ref{secBKAlgorithm} for the SVJ model. We recall that this algorithm also appeared in \cite{BroadieKa06} and in similar form as Algorithm 7.1 in \cite{KornKoKr10}. Further, since $\tilde{Y}_i$ follow the lognormal distribution with mean $\mu_S$ and variance $\sigma^2_S$, it is clear that
\begin{displaymath}
\sum^{N_t}_{j=1} \log(\tilde{Y}_j) \vert N_t \sim N \left( N_t \mu_s , N_t \sigma^2_S \right) \, .
\end{displaymath}

\begin{algorithm}
\caption{Exact simulation algorithm for the SVJ model}\label{algSVJEurop}
\begin{algorithmic}[1]
\STATE Simulate $V_t$ given $V_0$
\STATE Generate a sample from the distribution of $\int^t_0 V_s ds$ given $V_t$ and $V_0$
\STATE Recover $\int^t_0 \sqrt{V_s} dW^1_s $ from \eqref{eqvarHeston} given $V_t$, $V_0$ and $\int^t_0 V_s ds$
\STATE Generate $\tilde{S}_t$
\STATE Generate $N_t$
\STATE Generate $\prod^{N_t}_{j=1} \tilde{Y}_j$, given $N_t$
\end{algorithmic}
\end{algorithm}

There are alternative approaches to simulating $\prod^{N_t}_{j=1} \tilde{Y}_j$ as required in Algorithm~\ref{algSVJEurop}. As in Section 3.5 of \cite{Glasserman04}, one can simulate $N_t$ by simulating the jump times of the Poisson process. Furthermore, as discussed in \cite{BroadieKa06}, given $N_t$, one can simulate the jump sizes $\tilde{Y}_i$, $i=1, \dots, N_t$ individually. However, Algorithm~\ref{algSVJEurop} results in a problem that is of fixed dimension, in particular, the dimension of the problem in Algorithm~\ref{algSVJEurop} is five, i.e. five random uniforms (or a QMC point from $[0,1]^5$) are used to obtain a realization of $S_T$. Having a problem of fixed dimensionality is important when applying QMC methods, which is the ultimate goal of this paper, hence we choose the formulation presented in Algorithm~\ref{algSVJEurop}.

\subsection{Path-dependent options in the SVJ model} \label{subsecpathdepSVJmodel}

As in Section \ref{secBridgesamplingBK}, we now turn to the problem of studying the pricing of options whose payoff is a function of
\begin{displaymath}
S_{t_1}, \dots, S_{t_h} \, ,
\end{displaymath}
where we choose $h=2^m$ for simplicity. A naive approach, analogous to Algorithm~\ref{Naive}, is given by Algorithm~\ref{algnaiveBKSVJ}.

\begin{algorithm}
\caption{Naive version of the exact simulation algorithm for the SVJ model} \label{algnaiveBKSVJ}
\begin{algorithmic}[1]
\STATE Simulate $( V_{t_i} )^{h}_{i=1}$ in the following order
\begin{displaymath}
V_{t_1} , V_{t_2} , \cdots , V_{t_h}.
\end{displaymath}
\STATE Simulate $ \left( \int^{t_i}_{t_{i-1}} V_s ds \vert V_{t_{i-1}} , V_{t_i} \right)$ , $i=1, \dots , h$.
\STATE Simulate $(S_{t_i})^{N}_{i=1}$ in the following order
\begin{displaymath}
S_{t_1} , S_{t_2} , S_{t_3} , \cdots , S_{t_h} .
\end{displaymath}
\STATE Simulate $(N_{t_i})^{h}_{i=1}$ in the following order
\begin{displaymath}
N_{t_1}, N_{t_2} , \dots, N_{t_h}.
\end{displaymath}
\STATE Simulate $\left( \prod^{N_{t_i}}_{j=N_{t_{i-1}}+1} \tilde{Y}_j \vert N_{t_{i-1}}, N_{t_i}  \right)$, $i=1, \dots, h$, in the following order
\begin{displaymath}
\prod^{N_{t_1}}_{j=1} \tilde{Y}_j, \prod^{N_{t_2}}_{j=N_{t_{1}}+1} \tilde{Y}_j, \ldots, \prod^{N_{t_h}}_{j=N_{t_{h-1}}+1} \tilde{Y}_j.
\end{displaymath}
\end{algorithmic}
\end{algorithm}

\subsection{Bridge sampling for path-dependent options in the SVJ model}

Now, similar to the case of the Heston model presented in Section~\ref{secBridgesamplingBK}, we now propose an algorithm that allocates the early dimensions to variates with high variances. As such, we propose a bridge sampling algorithm for the SVJ model. We use the approach from Section \ref{secBridgesamplingBK} to deal with the diffusion component $( \tilde{S}_{t_i})^N_{i=1}$ and the approach proposed by Baldeaux in \cite{Bal08} to deal with the jump part. This results in Algorithm~\ref{algbridgeSVJBK}.

The next lemma, also presented as Lemma 3.1 in \cite{Bal08}, allows us to perform Step 4 of Algorithm \ref{algbridgeSVJBK} efficiently.

\begin{lemma} Let $s<u<t$ and $k_1 <k_2$. Then conditional on $N_s=k_1$ and $N_t=k_2$ the increment $N_u - N_s$ has the binomial distribution with parameters $k_2 - k_1$ and $(u-s)/(t-s)$.
\end{lemma}

Finally, the next lemma provides the tool required to complete the final step of Algorithm~\ref{algbridgeSVJBK}.

\begin{lemma} Given times $t_1 < t_2 < t_3$, the joint distribution of $$\left( \sum^{N_{t_1}}_{j=1} \log(\tilde{Y}_j) , \sum^{N_{t_2}}_{j=1} \log(\tilde{Y}_j) , \sum^{N_{t_3}}_{j=1} \log(\tilde{Y}_j) \right)$$ conditional on $\left(N_{t_1}, N_{t_2}, N_{t_3}\right)$, is given by
\begin{displaymath}
\left. \left( \begin{array}{c} \sum^{N_{t_1}}_{j=1} \log(\tilde{Y}_j) \\ \sum^{N_{t_2}}_{j=1} \log(\tilde{Y}_j) \\ \sum^{N_{t_3}}_{j=1} \log(\tilde{Y}_j) \end{array} \right) \right| N_{t_1}, N_{t_2}, N_{t_3} \sim N \left( \left[ \begin{array}{c} \mu_1 \\ \mu_2 \\ \mu_3 \end{array} \right] \, , \, \left[ \begin{array}{ccc} \sigma^2_1 & \sigma^2_1 & \sigma^2_1 \\ \sigma^2_1 & \sigma^2_2 & \sigma^2_s \\ \sigma^2_1 & \sigma^2_2 & \sigma^2_3 \end{array}  \right] \right) \, ,
\end{displaymath}
where $\mu_i = N_{t_i} \mu_S$ and $\sigma^2_i = N_{t_i} \sigma^2_S$. Further, we have that
\begin{equation*}
\left.\sum^{N_{t_2}}_{j=1} \log(\tilde{Y}_j) \right| \sum^{N_{t_1}}_{j=1} \log(\tilde{Y}_j) \, , \, \sum^{N_{t_3}}_{j=1} \log(\tilde{Y}_j) \sim N(M,\Sigma)
\end{equation*}
where
\begin{equation*}
M:=x_1 + m_2 - m_1 +\frac{\sigma^2_2 - \sigma^2_1}{\sigma^2_3 - \sigma^2_1} \left( x_3 - x_1 + m_1 - m_3 \right),
\end{equation*}
and
\begin{equation*}
\Sigma :=\sigma^2_2 - \sigma^2_1 - \frac{( \sigma^2_2 - \sigma^2_1 )^2}{\sigma^2_3 - \sigma^2_1}.
\end{equation*}
\end{lemma}

\begin{algorithm}
\caption{Bridge version of the exact simulation algorithm for the SVJ model}\label{algbridgeSVJBK}
\begin{algorithmic}[1]
\STATE Simulate $( V_{t_i} )^{N}_{i=1}$ in the following order
\begin{displaymath}
V_{t_N} \, , \, V_{t_{\frac{N}{2}}} \, , \,  V_{t_{\frac{N}{4}}} \, , \,  V_{t_{\frac{3 N}{4}}} \, , \cdots
\end{displaymath}
\STATE Simulate $ \left( \int^{t_i}_{t_{i-1}} V_s ds \vert V_{t_{i-1}} , V_{t_i} \right)$ , $i=1, \dots , N$.
\STATE Simulate $(\tilde{S}_{t_i})^{N}_{i=1}$ in the following order
\begin{displaymath}
\tilde{S}_{t_N} \, , \, \tilde{S}_{t_{\frac{N}{2}}} \, , \, \tilde{S}_{t_{\frac{N}{4}}} \, , \, \tilde{S}_{t_{\frac{3N}{4}}} \, , \, \dots
\end{displaymath}
\STATE Simulate $(N_{t_i})^{N}_{i=1}$ in the following order
\begin{displaymath}
N_{t_N}, N_{t_\frac{N}{2}} , N_{t_{\frac{N}{4}}}, N_{t_{ \frac{ 3 N}{4}}} , \dots
\end{displaymath}
\STATE Simulate $ ( \prod^{N_{t_i}}_{j=1} \tilde{Y}_j )^N_{i=1}$ conditional on $( N_{t_i} )^N_{i=1}$ in the following order
\begin{displaymath}
\prod^{N_{t_N}}_{j=1} \tilde{Y}_j \, , \, \prod^{N_{t_\frac{N}{2}}}_{j=1} \tilde{Y}_j \, \prod^{N_{t_\frac{N}{4}}}_{j=1} \tilde{Y}_j \, , \, \prod^{N_{t_\frac{3N}{4}}}_{j=1} \tilde{Y}_j
\end{displaymath}
\end{algorithmic}
\end{algorithm}

\subsection{Numerical results}

To demonstrate our bridging technique, we estimate the price of an Asian call option where the asset price process is given by a SVJ model. We compare the prices obtained with standard Monte Carlo, QMC using Sobol points and Owen's scrambling algorithm, and the bridge construction using Sobol points and Owen's scrambling. These results are presented in Table~\ref{TableAsianSVJ} and agree with the ones presented in Table \ref{TableAsian}: QMC methods improve on Monte Carlo methods, but the effectiveness of Monte Carlo methods can again be enhanced via bridge constructions.

\begin{table}
{\small
\caption{Standard errors of Asian option price where the asset price process is given by a SVJ model using Monte Carlo (M), Quasi-Monte Carlo (Q), and the Bridge Quasi-Monte Carlo (B) algorithms. These values are based on 30 independent batches and $m$ trials.}\label{TableAsianSVJ}
\begin{tabular}{c|cccccccccc}
\hline
$n$ & $m$: & $2^6$ & $2^7$ & $2^8$ & $2^9$ & $2^{10}$ & $2^{11}$ & $2^{12}$ & $2^{13}$ & $2^{14}$ \\
	\hline
\multirow{3}{*}{$4$} 
& M & 0.1354 & 0.0904 & 0.0717 & 0.0525 & 0.0254 & 0.0199 & 0.0152 & 0.0105 & 0.0065 \\
& Q & 0.0549 & 0.0401 & 0.0293 & 0.0116 & 0.0064 & 0.0055 & 0.0042 & 0.0026 & 0.0018 \\
& B & 0.0595 & 0.0403 & \hl 0.0217 & 0.0214 & 0.0171 & \hl 0.0041 & \hl 0.0032 & 0.0029 & \hl 0.0015 \\
\hline
\multirow{3}{*}{8}
& M & 0.1313 & 0.0873 & 0.0442 & 0.0421 & 0.0284 & 0.0200 & 0.0089 & 0.0078 & 0.0060 \\
& Q & 0.0488 & 0.0308 & 0.0304 & 0.0198 & 0.0188 & 0.0089 & 0.0031 & 0.0027 & 0.0016 \\
& B & \hl 0.0455 & \hl 0.0235 & \hl 0.0224 & \hl 0.0151 & \hl 0.0088 & \hl 0.0087 & 0.0044 & 0.0036 & 0.0027 \\
\hline
\multirow{3}{*}{16}
& M & 0.0786 & 0.0708 & 0.0447 & 0.0427 & 0.0278 & 0.0199 & 0.0127 & 0.0107 & 0.0060 \\
& Q & 0.0618 & 0.0415 & 0.0312 & 0.0185 & 0.0109 & 0.0088 & 0.0062 & 0.0035 & 0.0021 \\
& B & \hl 0.0520 & \hl 0.0349 & \hl 0.0196 & \hl 0.0173 & \hl 0.0100 & \hl 0.0069 & \hl 0.0058 & \hl 0.0023 & \hl 0.0015 \\
\hline
\multirow{3}{*}{32}
& M & 0.1218 & 0.0622 & 0.0508 & 0.0346 & 0.0265 & 0.0172 & 0.0116 & 0.0082 & 0.0060 \\ 
& Q & 0.0615 & 0.0318 & 0.0353 & 0.0213 & 0.0146 & 0.0084 & 0.0046 & 0.0037 & 0.0020 \\
& B & \hl 0.0359 & 0.0327 & \hl 0.0193 & \hl 0.0140 & \hl 0.0105 & \hl 0.0052 & \hl 0.0031 & \hl 0.0019 & \hl 0.0012 \\
\hline
\end{tabular}
}
\\[0.3em]
{\small Option parameters: $S=100$, $K=100$, $V_0=0.010201$, $\kappa=6.21$, $\theta=0.019$, $\sigma=0.61$, $\rho=-0.70$, $r=3.19\%$, $\lambda = 0.11$, $\mu_s = -0.1391$, and $\sigma_s = 0.15$, with $n$ time monitors over the time period $[0,1]$.}
\end{table}

\section{Further Extensions} \label{secFurExt}

We now propose a number of extensions to the results of the previous sections, in particular, the algorithms discussed in Section~\ref{secBridgesamplingBK} and Section~\ref{secExtensionSVJmodel} can be modified to solve some further problems of interest to practitioners in finance. First, we discuss how to compute ``greeks''. Second, we show how our algorithm can be enhanced for barrier options. Third, we provide an extension to the multidimensional and multi-asset setting. Fourth, we show how to price path-dependent options when the asset price process follows the $3/2$ model.

\subsection{Computation of greeks}

In a subsequent paper by Broadie and Kaya \cite{BroadieKa04}, it was shown how to modify their exact simulation algorithm to compute greeks. We now summarize their methodology and show how to adapt their method to our bridge algorithms for path-dependent options. To simplify the exposition, we focus on the Heston case, but we can also handle the SVJ case. We discuss the pathwise (PW) and the Likelihood Ratio (LR) method \cite{BroadieGl96,Glasserman04}. Using the notation from \cite{BroadieKa04}, we assume that the option price is given by
\begin{displaymath}
\alpha(\theta) = E \left[ f( \theta) \right] \, ,
\end{displaymath}
where $f$ denotes the discounted payoff function and $\theta$ the parameter of interest, i.e., we are interested in computing $\alpha'(\theta)$. Regarding the PW method, we have
\begin{displaymath}
\alpha'(\theta) = \frac{d}{d \theta} E \left[ f(\theta) \right] = E \left[ f'(\theta) \right] \, ,
\end{displaymath}
assuming the interchange of differentiation and integration is permitted. For the LR method, we consider $\theta$ as a parameter of the transition density of the random variable under consideration, say $X$. Denoting this density by $g_{\theta}(\bx)$, we have
\begin{displaymath}
\alpha'(\theta) = \frac{d}{d \theta} E \left[ f(X) \right] = \int_{\R^d} f(\bx) \frac{d}{d \theta} g_{\theta}(\bx) d\bx \, .
\end{displaymath}
Now, we rewrite this as
\begin{displaymath}
\alpha'(\theta) = \int_{\R^d} f(\bx) \frac{g'_{\theta}(\bx)}{g_{\theta}(\bx)} d \bx = E \left[ f(X) \frac{g'_{\theta}(X)}{g_{\theta}(X)} \right] \, .
\end{displaymath}
The quantity $\frac{g'_{\theta} (\bx)}{g_{\theta}(\bx)}$ is also known as the score function, and is of course independent of the particular payoff under consideration. Clearly, both approaches rely on the interchangeability of differentiation and expectation, and we refer the reader to \cite{BroadieGl96,Glasserman04} for details. To be able to apply the LR method, we need to have access to the density, denoted by $g_{\theta}(\bx)$ above. To achieve this, we apply a conditioning argument, from the law of iterated expectations,

\begin{equation} \label{eqcondexp}
E \left[ f(X) \right]=  E \left[ E \left[ f(X) \vert Y \right] \right],
\end{equation}
where $Y$ is a vector valued random variable. For the Heston model, $X$ will correspond to values of the stock price at discrete time intervals along a path, and $Y$ will be a set of state variables recording information about the variance path. To derive the LR and the PW estimator, Broadie and Kaya differentiate inside the expectation operator in \eqref{eqcondexp}. If the interchange of differentiation and integration is justified for the left hand side of \eqref{eqcondexp}, it is also justified for the right hand side. Though the PW approach can be applied to the left hand side of \eqref{eqcondexp}, in \cite{BroadieKa04} it is applied to the right hand side, to ensure the computational times are comparable.

To show how to apply the approach, we fix a partition of the time interval, $0=t_0<t_1< \dots < t_d=T$. As in the previous sections of the paper, we are interested in pricing a path-dependent option whose payoff is a function of the stock price vector $(S_{t_0}, \dots, S_{t_d})$. Assuming that we have simulated a path of the variance process using the first two steps of the algorithms in Sections \ref{secBridgesamplingBK} and \ref{secExtensionSVJmodel} and consider two consecutive times $t_i<t_j$. From Sections \ref{secBridgesamplingBK} and \ref{secExtensionSVJmodel}, we obtain values of $\int^{t_j}_{t_i} V_S ds$ and $\int^{t_j}_{t_i} \sqrt{V_s} dW^1_s$. Define
\begin{displaymath}
\bar{\sigma}^2_j := \frac{(1-\rho^2) \int^{t_j}_{t_i} V_s ds}{t_j - t_i} \, ,
\end{displaymath}
and
\begin{displaymath}
\xi_j := \exp \left( - \frac{\rho^2}{2} \int^{t_j}_{t_i} V_s ds + \rho \int^{t_j}_{t_i} \sqrt{V_s} dW^1_s \right) \, .
\end{displaymath}
Then given $S_{t_i}$, and the variance path, the value of $S_{t_j}$ can be expressed as
\begin{displaymath}
S_{t_j} = S_{t_i} \xi_j \exp \left( (r - \frac{\bar{\sigma}^2_j}{2} )(t_j - t_i) + \bar{\sigma}_j \sqrt{t_j - t_i} Z \right) \, ,
\end{displaymath}
where $Z$ is a standard normal random variable. Hence we take $Y$ in \eqref{eqcondexp} to be the variance path, and hence reduce the distribution in the inner expectation to a sequence of lognormal random variables, which is crucial to the LR method, as we need to be able to compute the score function. Again, as an example of a payoff, we consider again an Asian option with strike $K$, expiry $T$ and payoff
$\left( \frac{1}{d} \sum^d_{i=1} S_{t_i} - K \right)^+$.
Once the pathwise and likelihood ratio estimates are given, the way to employ the algorithm from Section \ref{secBridgesamplingBK} becomes obvious. As such, to simplify notation, we set
$\bar{S} = \frac{1}{d} \sum^d_{i=1} S_{t_i}$,
define the time increment as $\Delta t_i := t_i - t_{i-1}$, and pose
\begin{displaymath}
d_i = \frac{\left(  \log (S_{t_i} / (S_{t_{i-1}} \xi_i) ) - ( r - \frac{1}{2} \bar{\sigma}^2_i ) \Delta t_i  \right)}{\bar{\sigma}_i \sqrt{\Delta t_i}} \, ,
\end{displaymath}
where $\bar{\sigma}^2_i$ is the variance between $t_{i-1}$ and $t_i$. Then the pathwise and likelihood ratio estimates are as follows.

\subsubsection{Pathwise (PW) estimators}
\begin{eqnarray*}
\mbox{Delta } &:& e^{-r T} \1_{\bar{S} \geq K} \frac{\bar{S}}{S_{t_0}},
\\ \mbox{Rho } &:& e^{-r T} \1_{\bar{S} \geq K} \left( \frac{1}{d} \sum^d_{i=1} S_{t_i} t_i - T(\bar{S} - K) \right).
\end{eqnarray*}

\subsubsection{Likelihood ratio estimators}

\begin{eqnarray*}
\mbox{Delta } &:& e^{-r T} (\bar{S} - K)^+ \left( \frac{d_1}{S_{t_0} \bar{\sigma}_1 \sqrt{\Delta t_1} } \right),
\\ \mbox{Gamma } &:& e^{-r T} (\bar{S} - K)^+ \left( \frac{d^2_1 - d_1 \bar{\sigma}_1 \sqrt{\Delta t_1} -1}{S^2_{t_0}} \bar{\sigma}^2_1 \Delta t_1 \right),
\\ \mbox{Rho } &:& e^{-r T} ( \bar{S} - K)^+ \left( -T + \sum^d_{i=1} \frac{d_i \sqrt{\Delta t_i}}{\bar{\sigma}_i} \right).
\end{eqnarray*}

\subsection{Barrier options} \label{secGlassStaumBarrier}

Consider a barrier option with monitoring dates $(t_1 \, , \, t_2 \, , \, \dots \, , \, t_h)$, where we choose $h:=2^m$ for simplicity. If the option is not knocked out at maturity $T=t_h$, it pays off $f(S_T)$. Following the approach of Glasserman and Staum \cite{GlassermanSt01}, because of the knock-out feature, $(S_{t})_{t \ge 0}$ takes a value in $\R^+ \cup \Delta$, where $\Delta$ is an absorbing state. If $S$ crosses the barrier at time $t_i$, the option is knocked out, and for all $j \geq i$, $S_j= \Delta$. Define $A_i$ to be the indicator function $\mathbf{1} (S_{t_i} \neq \Delta)$, so $A_i=1$ means that the option is alive at time $t_i$. Assume the barrier is a price level $H<S_0$, so $A_i=1$ if the stock price has not crossed below the barrier $H$ by step $i$. A down-and-out call in this model has the discounted pay-off $A_m e^{- r T} (S_T - K)^+$, where $K$ is the strike price, $r$ the constant interest rate, $T=t_h$ denotes maturity. This allows us to propose a naive approach to pricing a barrier option in Algorithm~\ref{algnaiveBKbarrier}.

Sampling conditionally on one-step survival uses Algorithm~\ref{algnaiveBKbarrier} but in step $3$ we use \eqref{eqnaivebridgesampling} where $U = ( 1 - p(S_{t_i})) + V p(S_{t_i})$,
with $V$ uniformly distributed on $(0,1)$, and
\begin{eqnarray*}
p(S_{t_i}) &=&  P \left( S_{t_{i+1}} \geq H \bigl \vert \bigr. S_{t_i}, V_{t_{i+1}}, V_{t_i} , \textstyle{\int^{t_{i+1}}_{t_i} V_s ds} \right)
\\ &=& \Phi \left( \frac{ \log \left( \frac{S_{t_i}}{H} \right) + m(t_i, t_{i+1}) }{\sigma(t_i, t_{i+1})} \right).
\end{eqnarray*}
We refer the reader to Section 2.2 in \cite{GlassermanSt01} and in particular equations (11), (12), and (13).

\begin{algorithm}
\caption{Bridge version of the Broadie-Kaya exact simulation algorithm} \label{algnaiveBKbarrier}
\begin{algorithmic}[1]
\STATE Simulate $( V_{t_i} )^{h}_{i=1}$ as in Algorithm~\ref{Naive} or Algorithm~\ref{Bridge}
\STATE Simulate $ \left( \int^{t_i}_{t_{i-1}} V_s ds \vert V_{t_{i-1}} , V_{t_i} \right)$ for $i=1, \dots , h$.
\STATE Simulate $(S_{t_i})^{h}_{i=1}$ as follows: Given $S_{t_i} \, , \, V_{t_i} \, , \, V_{t_{i+1}}$, and $\int^{t_{i+1}}_{t_i} V_s ds$, set
\begin{equation} \label{eqnaivebridgesampling}
S_{t_{i+1}} \gets S_{t_i} \exp \left( m(t_i, t_{i+1}) + \sigma( t_i , t_{i+1}) \Phi^{-1} (U ) \right) \, ,
\end{equation}
where $m(u,t) := r(t-u) - \frac{1}{2} \int^t_u V_s ds + \rho \int^t_u \sqrt{V_s} dW^1_s $ and $\sigma^2(u,t) := ( 1- \rho^2) \int^t_u V_s ds$, where $U$ is uniformly distributed on $(0,1)$ and $\Phi$ denotes the standard normal cdf.
\end{algorithmic}
\end{algorithm}

\subsection{Multi-asset stochastic volatility models}

In this subsection, we show that the approach discussed in this paper can also be extended to a multi-asset stochastic volatility setting. When studying multidimensional stochastic volatility models, it is convenient to specify the model in such a way that the resulting multidimensional stochastic volatility process is an affine process \cite{DaiSIngleton00, Duffie00}. For affine stochastic volatility models, characteristic functions are known to satisfy a particular set of Ricatti equations, resulting in a tractable model in which path-independent European options can be efficiently priced using Fourier transforms \cite{carrmadan99}. As we now demonstrate, the methodology developed in this paper is particularly amenable to the affine structure of the model: To be precise, we introduce two stock price processes, $S^1$ and $S^2$, and three variance processes, $V^1$, $V^2$, and $V^3$. We model the covariation between the stock prices via the variance processes as
\begin{align*}
d S^1_t &= S^1_t \left(  r dt + \sqrt{V^1_t} d Z^1_t + \sqrt{V^3_t} d Z^3_t  \right),
\\ d S^2_t &= S^2_t \left( r dt + \sqrt{V^2_t} d Z^2_t + \sqrt{V^3_t} dZ^3_t \right),
\end{align*}
where for $i=1,2,3$ we have
\begin{equation}
d V^i_t = \kappa_i \left( \theta_i - V^i_t \right) dt + \sigma_i \sqrt{V^i_t} dW^i_t, \qquad Z^i_t = \rho_i W^i_t + \sqrt{1 - \rho^2_i} B^i_t,
\end{equation}
where $(W^1, W^2, W^3, B^1, B^2, B^3)$ is a standard six-dimensional Brownian motion. We point out that each stock price process corresponds to a Bi-Heston model, as introduced by Christoffersen, Heston, and Jacobs in \cite{Chis09}. The stochastic covariation between $S^1$ and $S^2$ is given by $\langle S^1 , S^2 \rangle_t = \int^t_0 S^1_s S^2_s V^3_s ds$. Using the terminology of \cite{DaiSIngleton00}, this model is an $A_3(5)$. As in Section \ref{secBKAlgorithm}, we use the following representation for stock prices:
\begin{align*}
S^i_t &= S^i_0 \exp \left( r t - \frac{1}{2} \int^t_0 V^i_s ds - \frac{1}{2} \int^t_0 V^3_s ds   \right)
\\ & \quad \times \exp \left(  \frac{\rho_i}{\sigma_i}  \left( V^i_t - V^i_0 + \int^t_0 \kappa_i V^i_s ds - \kappa_i \theta_i t \right) + \sqrt{1 - \rho^2_i} \int^t_0 \sqrt{V^i_s} d B^i_s  \right)
\\ & \quad \times \exp \left(  \frac{\rho_3}{\sigma_3}  \left( V^3_t - V^3_0 + \int^t_0 \kappa_3 V^3_s ds - \kappa_3 \theta_3 t \right) + \sqrt{1 - \rho^2_3} \int^t_0 \sqrt{V^3_s} d B^3_s  \right),
\end{align*}
where $i=1,2$. From this representation, it is now clear how to produce an algorithm which allows us to handle multiasset stochastic volatility models. We propose this as Algorithm~\ref{alg:multiasset}.

\begin{algorithm}
\caption{Exact Simulation Algorithm for the Multiasset model}\label{alg:multiasset}
\begin{algorithmic}[1]
\STATE Simulate $V^1_t, V^2_t, V^3_t$ using the
noncentral $\chi^2$-distribution
\STATE Simulate $\int^t_0 V^1_s ds \vert V^1_t \, , \, \int^t_0 V^2_s ds \vert V^2_t \, , \, \int^t_0 V^3_s ds \vert V^3_t$
\STATE For $i=1,2,3$; Compute
\begin{displaymath}
\int^t_0 \sqrt{V^i_t} d W^i_s = \frac{1}{\sigma_i} ( V^i_t - V^i_0 + \int^t_0 \kappa_i V^i_s ds - \kappa_i \theta_i t)
\end{displaymath}
\STATE Simulate $\log ( S^1_t) \sim N( \mu_1(t) , \tilde{\sigma}^2_1 (t) )$, where
\begin{align*}
\mu_1(t) &= \log(S^1_0) + r t - \frac{1}{2} \int^t_0 V^1_s ds - \frac{1}{2} \int^t_0 V^3_s ds + \rho_1 \int^t_0 \sqrt{V^1_s} dW^1_s + \rho_3 \int^t_0 \sqrt{V^3_t} d W^3_s
\\
\tilde{\sigma}^2_1 (t) &= (1 - \rho^2_1) \int^t_0 V^1_s ds + (1 - \rho^2_3) \int^t_0 V^3_s ds
\end{align*}
and $\log(S^2_t) \sim N( \mu_2(t) , \tilde{\sigma}^2_2 (t) )$, where
\begin{align*}
\mu_2(t) &= \log(S^2_0) + r t - \frac{1}{2} \int^t_0 V^2_s ds - \frac{1}{2} \int^t_0 V^3_s ds + \rho_2 \int^t_0 \sqrt{V^2_s} d W^2_s + \rho_3 \int^t_0 \sqrt{V^3_s} d W^3_s
\\ \tilde{\sigma}^2_2(t) &= (1 - \rho^2_2) \int^t_0 V^2_s ds + ( 1 - \rho^2_3) \int^t_0 V^3_s ds.
\end{align*}
\end{algorithmic}
\end{algorithm}

We point out that Algorithm~\ref{alg:multiasset} samples the joint distribution $(S^1_t, S^2_t)$. However, from Algorithm \ref{secBridgesamplingBK}, it is clear how to modify Algorithm \ref{alg:multiasset} to allow for path-dependent payoffs, i.e., how to sample the joint distribution
\begin{displaymath}
(S^1_{t_1}, \dots, S^1_{t_d}, S^2_{t_1}, \dots, S^2_{t_d}).
\end{displaymath}

\subsection{$3/2$ Model}

The $3/2$ model was introduced by Heston in \cite{Heston97}, see also \cite{CarrSu07, ItkinCa10, Lewis00}. Under the $3/2$ model, the stock price is given by
\begin{align*} \label{eqstock3over2}
\frac{d S_t}{S_t} &= r dt + \sqrt{V_t} \rho dW^1_t + \sqrt{V_t} \sqrt{1- \rho^2} dW^2_t,\\
dV_t &= \kappa V_t \left( \theta - V_t  \right) dt + \epsilon V^{3/2}_t d W^1_t.
\end{align*}
Recently, Baldeaux \cite{Bal12} showed how to modify the Broadie and Kaya approach from Section~\ref{secBKAlgorithm} to handle the $3/2$ model: First, we introduce the process $X_t=V_t^{-1}$, $t \geq 0$, whose dynamics are given by
\begin{equation} \label{eqXprocess3over2}
d X_t = \left( \kappa + \epsilon^2 - \kappa \theta X_t \right) dt - \epsilon \sqrt{X_t} d W^1_t \, .
\end{equation}
It is now easily verified that the stock price $S_u$ is given by
\begin{equation}
S_t \exp \left( r (u-t) - \tfrac{1}{2} \int^u_t \left( X_s \right)^{-1} ds + \rho \int^u_t \left( \sqrt{X_s} \right)^{-1} dW^1_s + \sqrt{1- \rho^2} \int^u_t \left( \sqrt{X_s} \right)^{-1} dW^2_s  \right) \, .
\end{equation}
Equation \eqref{eqXprocess3over2} shows that $X= \left\{X_t \, , \, t \geq 0 \right\}$ is a square-root process. We recall the algorithm presented in \cite{Bal12} for the exact simulation of the $3/2$ model in Algorithm~\ref{alg:3over2}.
\begin{algorithm}
\caption{Exact Simulation Algorithm for the $3/2$ model}\label{alg:3over2}
\begin{algorithmic}[1]
\STATE Simulate $X_u \vert X_t$ using the
noncentral $\chi^2$-distribution
\STATE Simulate
$\int^u_t \frac{ds}{X_s} \vert X_t , X_u$
\STATE Recover $\int^u_t \left( \sqrt{X_s} \right)^{-1} dW^1_s$ from
\begin{equation*} \label{eqBMintegral}
\int^u_t \left( \sqrt{X_s} \right)^{-1} dW^1_s = \frac{1}{\epsilon} \left( \log \left( \frac{X_t}{X_u} \right) + \left( \kappa + \frac{\epsilon^2}{2} \right) \int^u_t \frac{ds}{X_s}   - \kappa \theta (u - t) \right).
\end{equation*}
\STATE Simulate $S_u$ given
$S_t$, $\int^u_t \left( \sqrt{X_s} \right)^{-1} dW^1_s$ and
$\int^u_t \left( X_s \right)^{-1} ds$ via
\begin{equation*}
\log (S_u) \sim N \left( \log (S_t) + r(u-t)  - \frac{1}{2} \int^u_t \left( X_s \right)^{-1} ds + \rho \int^u_t \left( \sqrt{X_s} \right)^{-1} dW^1_s , \sigma^2(t,u) \right),
\end{equation*}
where
    \begin{displaymath}
    \sigma^2(t,u) = \left( 1 - \rho^2 \right) \int^u_t X^{-1}_s ds  \, .
    \end{displaymath}
\end{algorithmic}
\end{algorithm}
As for the Heston model, Step 2 is the difficult step and, as in Section~\ref{secBKAlgorithm}, one proceeds by computing the relevant Laplace transform \cite{Bal12}. We can use the bridge constructions for the square-root process and the Brownian bridge from Section~\ref{secBridgesamplingBK} for steps 1 and 3 of Algorithm~\ref{alg:3over2} respectively, and handle Step 2 as in \cite{Bal12}. Consequently, we can use the techniques from Section~\ref{secBridgesamplingBK} to also handle the $3/2$ model for path-dependent options.


\end{document}